# The Massive Fermion Phase for the U(N) Chern-Simons Gauge Theory in D=3 at Large N

by

William A. Bardeen
*Fermilab, MS 106, P.O. Box 500, Batavia, IL 60510*
bardeen(at)fnal.gov

## Abstract

We explore the phase structure of fermions in the U(N) Chern-Simons Gauge theory in three dimensions using the large N limit where N is the number of colors and the fermions are taken to be in the fundamental representation of the U(N) gauge group. In the large N limit, the theory retains its classical conformal behavior and considerable attention has been paid to possible AdS/CFT dualities of the theory in the conformal phase. In this paper we present a solution for the massive phase of the fermion theory that is exact to the leading order of 't Hooft's large N expansion. We present evidence for the spontaneous breaking of the exact scale symmetry and analyze the properties of the dilaton that appears as the Goldstone boson of scale symmetry breaking.

# 1. Introduction.

The most direct analog of quantum chromodynamics that describes the strong force in four dimensions is the Chern-Simons gauge theory in three dimensions. Both theories are classically conformal and both are thought to have simplifications in the 't Hooft large N limit [1] where N is the number colors of the nonabelian gauge group. At leading order in the large N expansion, the Chern-Simons theory retains its conformality and has been the subject of a number of interesting studies of the conformal phase for these theories and their possible dual descriptions as higher spin gravity theories on $AdS_4$ [2,3,4,5,6,7]. The U(N) Chern-Simons gauge theory with scalar quarks has recently been shown have a massive phase for certain critical combinations of the gauge and scalar coupling constants where the conformal symmetry is spontaneously broken [8]. A massless, U(N) singlet, scalar bound state is discovered by studying the gauge invariant correlation functions of the color singlet, scalar current densities. Its properties are consistent with those expected of the dilaton, the Goldstone boson of the spontaneous breaking of the scale symmetry.

In this paper we turn to the study of the quarks as fermions in the fundamental representation of the U(N) Chern-Simons gauge group. We explore the nature of a massive phase of the theory analogous to the previous study of scalar quarks. We obtain exact solutions for the gap equations for the quark self-energies, the scalar current vertex functions and the two-point scalar current correlator. With these solutions we explore the nature of the spontaneous symmetry breaking and the role of the dilaton in the infrared dynamics of the theory.

# 2. Chern-Simons Fermions. Gap Equations.

In three dimensions, Chern-Simons gauge theory is a topological field theory, and there are no actual degrees of freedom associated with the gluon dynamics. However, the Chern-Simons interactions do have important implications for the matter fields and their dynamics. We focus our study on gauge singlet operators, particularly the scalar current density.

The U(N) Chern-Simons Euclidean action is given by

$$S_{CS} = \frac{i\kappa}{8\pi} \varepsilon_{\mu\nu\rho} \int d^3x \, tr\left\{ A_\mu^a \partial_\nu A_\rho^a + i\frac{2}{3} A_\mu^a A_\nu^b A_\rho^c f^{abc} \right\} \tag{2.1}$$

where $\kappa$ is the Chern-Simons level. The quarks are described by a two component fermion field in three dimensons. The fermion action is given by

$$S_{fermion} = \int d^3x \left[ \psi^+ (\gamma_\mu D_\mu + m) \psi \right], \quad D_\mu = \partial_\mu + iA_\mu^a (T^a) \tag{2.2}$$

where $T^a = \lambda^a/2$ are the generators in the fundamental representation of U(N) and $\gamma_\mu = \sigma_\mu$, are the Pauli spin matrices.

Following previous work, we use a Euclidean light-cone gauge [2] that greatly simplifies the calculations and is essential to obtaining the results in this paper. In this gauge, $A_- = (A_1 - iA_2)/\sqrt{2} = 0$, the nonabelian terms in the gauge field action vanish. The calculation is further simplified by using the large N expansion where the leading order is described only by the sum of planar diagrams, and the color degrees of freedom contribute appropriate factors of N to the various diagrams. In light-cone gauge the gluon propagator is given by

$$G_{+3}(p) = -G_{3+}(p) = \frac{4\pi i}{\kappa}\frac{1}{p_-} = 4\pi i \frac{\lambda}{N}\frac{1}{p_-}, \quad \lambda \equiv \frac{N}{\kappa} \tag{2.3}$$

In the large N expansion, the gauge coupling, $\lambda$, is held fixed as N and $\kappa$ are taken to infinity. The large N expansion selects the planar diagram contributions to leading order for each process. Hence, the fermion self-energies are determined by the sum of the "rainbow" diagram contributions to the gap equation. The vertex functions require the sum of the "ladder" diagram contributions where the kernel is provided by the interaction in Eq.(2.3).

The algebra is simplified by using the following relations in light-cone coordinates

$$\frac{\partial}{\partial p_+}\frac{1}{(p-q)_-} = 2\pi\delta^2(p-q), \quad \frac{\partial}{\partial p_+}(p_-) = 0, \quad \frac{\partial}{\partial p_+}p_s = \frac{p_-}{p_s}, \quad p_s^2 = 2p_+p_- \tag{2.4}$$

We can now proceed to compute the fermion self-energy function in the light-cone gauge formalism. Because we are using a noncovariant gauge, the fermion self-energy function can have noncovariant terms. The inverse fermion propagator is given by

$$S^{-1}(p) = (i\gamma_\mu p_\mu + \Sigma(p)), \quad \Sigma(p) = \gamma_-\Sigma_+(p) + \Sigma_o(p)$$

$$S(p) = (-i\gamma_\mu p_\mu - \gamma_-\Sigma_+(p) + \Sigma_o(p))/D(p) \tag{2.5}$$

$$D(p) = (p_3^2 + 2p_+p_- - 2ip_-\Sigma_+(p) + \Sigma_o^2(p)) = (p_3^2 + p_s^2 + M^2)$$

The other noncovariant components of the self-energy function vanish in this gauge.

The gap equation for the self-energy function becomes

$$\Sigma(p) = m - 2\pi\lambda i \int \frac{d^3q}{(2\pi)^3} \frac{1}{(p-q)_-} \{\gamma_3 S(q)\gamma_- - \gamma_- S(q)\gamma_3\} \qquad (2.6)$$

where we have included a possible fermion mass term so we can tune to the critical point, if necessary. Because of the planar structure of the large N limit, Eq.(2.6) is exact to leading order in the large N expansion. From the form of the self-energy equation, the self-energies can only depend on the transverse momentum, $p_\pm$, and not $p_3$.

Using the simple form of the massive quark propagator in E.(2.5), we can immediately integrate the equation determining singlet self-energy, $\Sigma_o(p)$.

$$\Sigma_o(p) = m - 4\pi\lambda i \int \frac{d^3q}{(2\pi)^3} \frac{1}{(p-q)_-} \frac{iq_-}{q_3^2 + q_s^2 + M^2} = m - 4\pi\lambda i \int \frac{\pi d^2 q}{(2\pi)^3} \frac{1}{(p-q)_-} \frac{iq_-}{\sqrt{q_s^2 + M^2}}$$

$$= m - \frac{\lambda}{2}\int_{p_s^2}^{\Lambda^2} dq_s^2 \frac{1}{\sqrt{q_s^2 + M^2}} = m + \lambda\left(\sqrt{p_s^2 + M^2} - \Lambda\right) = m + \lambda\sqrt{p_s^2 + M^2} \qquad (2.7)$$

The linear divergence is discarded due to the conformal structure of the theory in the large N limit.

The noncovariant self-energy, $\Sigma_+(p)$, can be directly obtained from a consistency condition that follows from the construction of the quark propagator in Eq.(2.5) and our solution for $\Sigma_o(p)$ in Eq.(2.7).

$$2ip_-\Sigma_+(p) = \Sigma_o^2(p) - M^2 = \lambda^2 p_s^2 + 2\lambda m\sqrt{p_s^2 + M^2} + m^2 - (1-\lambda^2)M^2 \qquad (2.9)$$

However, $\Sigma_+(p)$ should also satisfy an integral equation following from Eq. (2.6).

$$\Sigma_+(p) = -4\pi\lambda i \int \frac{d^3q}{(2\pi)^3} \frac{1}{(p-q)_-} \frac{\Sigma_o(q)}{q_3^2 + q_s^2 + M^2} = -4\pi\lambda i \int \frac{\pi d^2 q}{(2\pi)^3} \frac{1}{(p-q)_-} \frac{\Sigma_o(q)}{\sqrt{q_s^2 + M^2}}$$

$$= -\frac{i}{p_-}\frac{\lambda}{2}\int_0^{p_s^2} dq^2 \frac{\Sigma_o(q)}{\sqrt{q_s^2 + M^2}} = -\frac{i}{p_-}\left(\frac{\lambda^2}{2} p_s^2 + \lambda m\left(\sqrt{p_s^2 + M^2} - M\right)\right) \qquad (2.10)$$

The two expressions for $\Sigma_+(p)$ differ by the constant term at $p_s^2 = 0$ in Eq.(2.9). In this paper, I will assume that Eq.(2.9) is the correct expression to use for the analysis of the massive quark phase. This implies there must be an additional contribution to Eq.(2.10). This contribution could be due to the macroscopic occupation of a zero mode at $q_s^2 = 0$ in Eq.(2.10). The zero mode contribution is not arbitrary but must have the specific value specified by the constant term in Eq.(2.9).

$$\Sigma_+(p) = -\frac{i}{p_-}\left\{R_+ + \left(\frac{\lambda^2}{2}p_s^2 + \lambda m\left(\sqrt{p_s^2 + M^2} - M\right)\right)\right\}, \quad 2R_+ = (m + \lambda M)^2 - M^2 \quad (2.11)$$

where $R_+$ arises from the zero mode contribution. We note that the zero mode does not contribute the integral equation that determines the singlet self-energy function, $\Sigma_o(p)$, as the integrand in Eq.(2.7) vanishes at $q_- = 0$. The presence of this zero mode may be a particular artifact associated with the use of light-cone gauge.

## 3. Vertex equations.

We now consider the scalar vertex function for the color singlet operator, $J_0(x) = \vec{\psi}^+\vec{\psi}(x)$. The equations for the scalar vertex functions simplify in the large N limit as only the ladder diagrams survive in light-cone gauge in the planar limit. The Schwinger-Dyson equations for the full scalar vertex take the form

$$V(p,k_3) = 1 + 2\pi\lambda i \int \frac{d^3q}{(2\pi)^3} \frac{1}{(p-q)_-}$$
$$\cdot \left\{\gamma_3 S_F(q)V(q,k_3)S_F(q+k_3)\gamma_- - \gamma_- S_F(q)V(q,k_3)S_F(q+k_3)\gamma_3\right\} \quad (3.1)$$

As for the self-energy function, noncovariant terms are generated as corrections for the vertex functions and $V(p,k_3) = V_0(p,k_3) + \gamma_- V_+(p,k_3)$ where the momentum transfer is taken to be nonzero only in the longitudinal direction, $k_3$. The integral equations for the components become

$$V_0(p,k_3) = 1 + 2\pi\lambda i \int \frac{d^3q}{(2\pi)^3} \frac{1}{(p-q)_-} \frac{1}{(q_3^2 + q_s^2 + M^2)} \frac{1}{\left((q_3 + k_3)^2 + q_s^2 + M^2\right)}$$
$$\cdot \left\{[2k_3 q_- + 4iq_-\Sigma_o(q)]V_0(q,k_3) + [4q_-^2]V_+(q,k_3)\right\} \quad (3.2)$$

$$V_+(p,k_3) = 2\pi\lambda i \int \frac{d^3q}{(2\pi)^3} \frac{1}{(p-q)_-} \frac{1}{(q_3^2 + q_s^2 + M^2)} \frac{1}{\left((q_3 + k_3)^2 + q_s^2 + M^2\right)}$$
$$\cdot \left\{[-2q_3(q_3 + k_3) - 4q_+q_- + 4iq_-\Sigma_+(q) + 2\Sigma_o^2(q)]V_0(q,k_3) + [2q_-k_3 - 4iq_-\Sigma_o(q)]V_+(q,k_3)\right\} \quad (3.3)$$

These integral equations can be solved exactly in the form

$$V_0(p,k_3) = A(k_3) + B(k_3)\Phi(p,k_3)$$

$$ip_- V_+(p,k_3) = (\Sigma_o(p) - ik_3/2)A(k_3) + B(k_3)(\Sigma_o(p) + ik_3/2)\Phi(p,k_3) \tag{3.4}$$

$$\Sigma_o(p)V_0(p,k_3) - ip_- V_+(p,k_3) = i(k_3/2)(A(k_3) - B(k_3)\Phi(p,k_3))$$

with

$$\Phi(p,k_3) = \exp(-2i\delta(p,k_3))$$

$$\delta(p,k_3) = \frac{1}{2}\lambda k_3 \int dx \left[x(1-x)k_3^2 + p_s^2 + M^2\right]^{-1/2} = \lambda \arctan\left(k_3/2\sqrt{p_s^2 + M^2}\right) \tag{3.5}$$

The vertex functions satisfy the boundary conditions

$$p \to \infty,\ V_0(p,k_3) \to 1;\quad p \to 0,\ ip_- V_+(p,k_3) \to 0 \tag{3.6}$$

that are used to determine the A and B coefficients.

$$1 = A(k_3) + B(k_3)$$

$$0 = (m + \lambda M - ik_3/2)\exp(i\delta)A(k_3) + B(k_3)(m + \lambda M + ik_3/2)\exp(-i\delta) \tag{3.7}$$

with $\delta = \delta(k_3) = \delta(0,k_3) = \lambda \arctan(k_3/2M)$ and

$$(A(k_3), B(k_3)) = \frac{1}{2}(+,-)\frac{1}{2}i\frac{1 + (k_3/2(m + \lambda M))\tan\delta}{\tan\delta - (k_3/2(m + \lambda M))} \tag{3.8}$$

These A and B coefficients can be inserted into the into Eq.(3.4) to obtain the complete solution for the scalar current vertex functions. We have retained the separate dependence on the bare mass, m, and the quark mass, M. We will see that the scale invariant limit for the massive phase requires a particular value for the ratio of these masses.

### Section 4. The scalar current two point function.

We are now in a position to compute the exact two-point correlation function of the scalar currents. This correlator is determined by a bubble loop integral involving one full vertex, the two full fermion propagators and one bare vertex.

$$\langle J_0(k_3)J_0(-k_3)\rangle/N = -\int \frac{d^3q}{(2\pi)^3} tr\{S(q)V(q,k_3)S(q+k_3)\} \tag{4.1}$$

This integral can be obtained directly from the vertex functions using their definition in Eq.(3.1).

$$\langle J_0(k_3) J_0(-k_3) \rangle / N = \frac{1}{(2\pi)^2 \lambda i} \int d^2 p\, \partial_{p_+}\left(\frac{1}{2} tr\{\gamma_+ V(p,k_3)\}\right)$$
$$= \frac{1}{(2\pi)^2 \lambda i} \int d^2 p\, \partial_{p_+} V_+(p,k_3) \quad (4.2)$$

The integrand in Eq.(4.2) is a total derivative and using the boundary conditions for the $V_+$ vertex in Eq. (3.6) we obtain

$$\langle J_0(k_3) J_0(-k_3) \rangle / N = -\frac{1}{2\pi\lambda}\left(ip_- V_+(p,k_3)\right),\ p_s^2 \to \infty \quad (4.3)$$

Given our solutions for the vertex functions, we take the appropriate limit in Eq.(3.4) to get our result for the scalar current correlator.

$$\langle J_0(k_3) J_0(-k_3) \rangle / N = -\frac{1}{2\pi}\Lambda + \frac{\lambda}{2\pi} i(k_3/2) - \frac{1}{2\pi}\frac{1}{\lambda} m$$
$$- \frac{1}{2\pi}\left(\frac{1}{\lambda} - \lambda\right)(k_3/2)\frac{1+(k_3/2(m+\lambda M))\tan(\lambda \arctan(k_3/2M))}{\tan(\lambda \arctan(k_3/2M)) - (k_3/2(m+\lambda M))} \quad (4.4)$$

It is remarkable that we are able to obtain an exact solution for the scalar current correlator for arbitrary values of the induced mass, M, and the bare mass parameter, m.

However, Eq.(4.4) retains some artifacts of our use of a noncovariant gauge for the calculation. These artifacts are contained in the analytic terms in Eq.(4.4). Our solution contains a linear divergence that is independent of the coupling constant and should be discarded due the conformal invariance of the theory at large N. The pure imaginary term is linear in the gauge coupling, $\lambda$, and the longitudinal momentum, $k_3$ should involve an explicit factor of the antisymmetric pseudotensor used to define the kinetic term for the Chern-Simons action. However, this term does not preserve the D=3 rotational invariance and must be discarded as an artifact of using the noncovariant light-cone gauge.

The Lorentz invariant and gauge invariant scalar correlator is then given solely by the final two terms in Eq.(4.4),

$$\langle J_0(\vec{k})J_0(-\vec{k})\rangle/N = -\frac{1}{2\pi}\frac{1}{\lambda}m$$
$$-\frac{1}{2\pi}\left(\frac{1}{\lambda}-\lambda\right)\left(\sqrt{\vec{k}^2}/2\right)\frac{1+\left(\sqrt{\vec{k}^2}/2(m+\lambda M)\right)\tan\left(\lambda\arctan\left(\sqrt{\vec{k}^2}/2M\right)\right)}{\tan\left(\lambda\arctan\left(\sqrt{\vec{k}^2}/2M\right)\right)-\left(\sqrt{\vec{k}^2}/2(m+\lambda M)\right)} \quad (4.5)$$

that should now be viewed as a function of the invariant, $\vec{k}^2$, and not simply $k_3^2$. As yet, we have not determined the appropriate value of the bare fermion mass parameter, m. We will see that the conformal limit requires a particular choice of the bare mass parameter, $m+\lambda M = M/\lambda$.

We can check the behavior in the conformal phase by taking the large momentum limit of the scalar correlator in Eq.(4.5). We find the result

$$\langle J_0(\vec{k})J_0(-\vec{k})\rangle/N = \frac{1}{4\pi}\left(\frac{1}{\lambda}-\lambda\right)\sqrt{\vec{k}^2}\tan\left(\lambda\frac{\pi}{2}\right), \quad \lambda < 2 \quad (4.6)$$

The correlator positive for the entire range of gauge couplings, $(0 < \lambda < 2)$, as expected for a consistent solution in Euclidean space. The next order correction to Eq.(4.6) is a constant term that depends only on the bare mass parameter, m, and the coupling constant, $\lambda$.

$$\langle J_0(\vec{k})J_0(-\vec{k})\rangle_{NLO} = \frac{\lambda}{2\pi}m\left\{-1+\left(\frac{1}{\lambda^2}-1\right)\tan^2\left(\lambda\frac{\pi}{2}\right)\right\} \quad (4.7)$$

We remark that the large momentum expansion of the scalar correlator has an analytic expansion in powers of the gauge coupling constant and the bare mass parameter, m.

Aside from some different sign conventions, the result presented in Eq.(4.6) agrees with calculations done strictly in the conformal phase of the theory except for the prefactor $(1/\lambda - \lambda)$ which was found instead to be $(1/\lambda)$ in other work [2,3,4,5,9]. This difference could have important consequences for the interpretation of the physics of the conformal phase including the critical endpoint of the theory. We discuss the origin of these differences in Appendix A where evidence is given for the validity of the result presented above.

**Section 5. Properties of the Massive Phase.**

We explore our solution for the massive phase U(N) Chern-Simons theory with quarks as fermions in the fundamental representation. We have computed the theory as a function of the induced quark mass, M, and a bare quark mass parameter, m. If the conformal symmetry of the original theory is to be preserved, these two masses should not be independent parameters. Furthermore, the spontaneous breaking of the scale symmetry should result in the appearance of a massless scalar Goldstone boson, the dilaton. We can use our solution to see if the dilaton pole emerges for a particular value of the bare mass, m, for fixed quark mass, M.

The dilaton pole can be seen as a singularity in low momentum behavior of the scalar current correlator. At zero momentum, the scalar correlator is given by

$$\langle J_0(k_3)J_0(-k_3)\rangle/N \to -\frac{1}{2\pi}\left(\frac{1}{\lambda}-\lambda\right)\frac{1}{(\lambda/M - 1/(m+\lambda M))}$$
$$= \frac{1}{2\pi}\left(\frac{1}{\lambda^2}-1\right)\frac{M(m+\lambda M)}{(M(1/\lambda - \lambda) - m)} \tag{5.1}$$

It is clear from Eq.(5.1) that a singularity can only occur when the condition $m = M/\lambda - \lambda M$ is satisfied. This requires a nonperturbative condition on the bare mass parameter. We will see that imposing this condition leads to a consistent picture of spontaneous scale symmetry breaking.

We now impose this condition on the bare mass parameter and make a more careful examination of the singular behavior of the correlator. Since the leading term in the denominator now vanishes, we must keep the higher order terms.

$$\langle J_0(\vec{k})J_0(-\vec{k})\rangle/N \to \frac{3}{2\pi}\frac{1}{\lambda^2}\frac{4M^3}{\vec{k}^2} + finite, \quad \vec{k}\to 0, \; m = M(1/\lambda - \lambda) \tag{5.2}$$

We, indeed, see the expected dilaton pole and its residue determines the dilaton "decay" constant, $f_D$.

$$J_0(x) = \vec{\psi}^+\vec{\psi}(x) = f_D D(x), \quad f_D^2 = \frac{6N}{\pi}\frac{M^3}{\lambda^2} \tag{5.3}$$

The dilaton decay constant is singular in the weak coupling gauge theory, $\lambda \to 0$, and the effective field theory for the dilaton becomes even more weakly coupled in this limit. The dilation coupling constants scale as $1/f_D$.

The behavior we observe Eq.(5.3) is in contrast to the scalar quark theory in [8] where the dilaton decay constant is singular as $\lambda \to 1$, the upper limit for the gauge coupling constant for scalar quarks. We recall that the analogous decay constant for the bosonic scalar current [8] is given by

$$J_{scalar}(x) = \vec{\phi}^+\vec{\phi}(x) = f_{scalar}D(x), \quad f_{scalar}^2 = \frac{3N}{2\pi}\frac{M}{(1-\lambda^2)} \tag{5.4}$$

The fermion theory appears to have a different critical point for the gauge coupling constant as the correlation function in Eq.(4.5) is consistent over the entire range $(0 < \lambda < 2)$ in contrast to the range $(0 < \lambda < 1)$ for the scalar quark theory. However, the result is in disagreement with other calculations [2,3,4,5,9] that find the critical point to be consistent with $\lambda = 1$, see Apendix A.

We can determine the dilaton correlation length when there is explicit symmetry breaking and the bare mass parameter differs from its critical value. The low momentum behavior will now be described by a massive dilaton with

$$\langle J_0(\vec{k})J_0(-\vec{k})\rangle \rightarrow \frac{f_D^2}{\vec{k}^2 + \mu^2}, \quad (\vec{k}^2, \mu^2) \ll M^2 \tag{5.5}$$

The value of dilaton mass, $\mu^2$, can be determined by combining the results of Eq.(5.1) and Eq.(5.2)

$$\mu^2 = \left(1/(m+\lambda M) - \lambda/M\right)\frac{12M^3}{\lambda(1-\lambda^2)}$$
$$\rightarrow \left(M(1/\lambda - \lambda) - m\right)12M/\left(\frac{1}{\lambda} - \lambda\right), \quad m \rightarrow M\left(\frac{1}{\lambda} - \lambda\right) \tag{5.6}$$

For the massive phase to exist, $\mu^2$ must be positive. Hence there are two regions where symmetry breaking can occur. For weak coupling, $0 < \lambda < 1$, the bare mass must be smaller than its critical value while for strong gauge coupling, $1 < \lambda < 2$, the bare mass must be larger than its critical value. Of course, the dilaton is massless for the entire range of gauge couplings at the critical point for the bare mass parameter reflecting the underlying scale invariance of the gauge theory.

The form factors for the coupling of the dilaton to the quarks to can also be found by computing the residue of the pole terms in the vertex functions in Eq.(3.4) to Eq.(3.8). These form factors are not gauge invariant since the quarks are not gauge singlets.

$$g_D = g_{D0} + \gamma_- g_{D+}$$

$$g_{D0} = \frac{1}{f_D}\frac{12M^2}{1-\lambda^2}\frac{M}{\sqrt{p_s^2 + M^2}} = \sqrt{\frac{24\pi}{N}}\frac{\lambda}{1-\lambda^2}\frac{M^{3/2}}{\sqrt{p_s^2 + M^2}}$$

$$g_{D+} = \frac{1}{f_D}(ip_+)\frac{12M}{\lambda}\frac{M}{\left(\sqrt{p_s^2 + M^2}\right)\left(\sqrt{p_s^2 + M^2} + M\right)}\frac{2M}{} \quad (5.7)$$

$$= (ip_+)\sqrt{\frac{24\pi}{NM}}\frac{M}{\left(\sqrt{p_s^2 + M^2}\right)}\frac{2M}{\left(\sqrt{p_s^2 + M^2} + M\right)}$$

These form factors fall off at high transverse momentum as one might expect for a low energy boundstate.

## 6. The Scalar Current Correlator.

We now discuss the general features of our solution for the two-point scalar current correlator. Because we are in the massive phase for the quark, the dilaton pole dominates the low momentum behavior of the correlator. The residue of the pole determines the dilaton decay constant which is found to scale like the inverse of the of the gauge coupling constant, Eq.(5.3). The high momentum behavior of the scalar correlator is described by the conformal behavior as shown in Eq.(4.6) and scales linearly with the momentum. We illustrate these behaviors in Figures 1-3.

We note that since dilation coupling to matter typically scales like the inverse of the decay constant, the dilaton tends to decouple for weak gauge coupling. However, we can see by the explicit calculation of the dilation-quark coupling form factors in Eq.(5.7) that this is compensated by an inverse power of the gauge coupling in the form factor and the dilation does not decouple as $\lambda \to 0$.

We consider the weak gauge coupling case in Figure 1 where we take $\lambda = 0.01$. Setting the scale breaking parameter to zero in Figure 1.a., we see the massless dilaton pole at low momentum and the conformal behavior at high momentum. The flat region in the middle illustrates the hierarchy of mass scales that exists at weak gauge coupling, the quark mass, M=1 in the Figures, and the bare quark mass $m = M(1/\lambda - \lambda)$. In Figure 1.b. we have adjusted the bare quark mass to allow for explicit symmetry breaking. The dilaton is clearly still present but has now acquired a mass through the explicit symmetry breaking in accordance with Eq.(5.5) and Eq.(5.6).

We next consider the case of moderately strong gauge coupling, $\lambda = 0.5$. As shown in Figures 2.a. and 2.b., there is now an immediate transition between the low momentum dilaton behavior and the high momentum conformal behavior. There is no hierarchy of scales as the gauge coupling is of order one.

We finally consider the case near the boundary for the gauge coupling at $\lambda = 2$. According to Eq.(4.6) the coefficient of the conformal term vanishes at the boundary and the theory becomes degenerate. In Figures 3.a and 3.b, we have chosen $\lambda = 1.999$ which is very close to the boundary. We again see a flat region between the expected infrared and ultraviolet regions. In this case, there is no mass hierarchy but the suppression of the conformal term delays its appearance in the correlator.

The massive phase also seems to exist at the boundary coupling. The dilaton still contributes a pole in the region of momentum below the dynamical quark mass, M, but the correlator is a constant for momenta larger than the quark mass. The theory still seems to retain a nontrivial infrared dynamics.

### 7. The Effective Action for the Dilaton.

The expression for the two-current correlator in Eq.(4.5) can be used to study the effective action for the dilaton valid as a low energy expansion. We can use the dependence on the bare quark mass to compute higher point correlation functions. Each derivative with respect to the bare quark mass will add an additional scalar current to the correlation function. These additional currents will all be evaluated at zero momentum while the two explicit currents have finite momentum. With our knowledge of these higher order correlation functions we will be able to infer the structure of the dilaton effective action. For this procedure to work we must take the limit where the dilaton mass is small but finite. This mass will appear as singularities of the correlation function indicating the presence the dilaton poles. The residue of these poles will determine the effective action.

We first will use this method to compute the nonderivative effective potential for the dilaton. We have determined the zero momentum correlator in Eq.(5.1) and it has a single pole as a function of the bare mass parameter, m. Each derivative with respect to the bare mass parameter adds at most a single power of the singularity that can be identified with the dilaton poles in the correlation function. The n-point correlation function requires (n-2) derivatives and it is easily seen that there can be only (n-1) powers of the inverse of the dilaton mass. However, the tree diagrams for the effective action contribution to the correlator will require (n) powers of the inverse dilaton mass singularity. Hence, we can conclude that there are no nonderivative contributions to the effective potential for the dilation. Of course, this is to be expected if the dilaton is identified as the Goldstone boson of a spontaneously broken scale symmetry.

However, the form of the correlator in Eq.(4.5) does allow for an effective action that is second order in the derivative expansion. To see this behavior, we observe that the denominator in Eq.(4.5) has terms that are higher order in the momentum, $\vec{k}$. If we expand a single power of $\vec{k}^2$ from the denominator we create an additional inverse power of the dilaton mass singularity. The power counting of the singularities now corresponds the presence of a nontrivial effective action and the factor of $\vec{k}^2$ implies the effective action is, at least, second order in the derivative expansion.

We need the second term in the low momentum expansion for the correlator in Eq.(4.5)

$$\langle J_0(\vec{k})J_0(-\vec{k})\rangle/N \to -\frac{1}{2\pi}\left(\frac{1}{\lambda}-\lambda\right)\frac{1}{\left(\frac{\lambda}{M}-\frac{1}{m+\lambda M}-\frac{1}{3}\lambda(1-\lambda^2)\frac{\vec{k}^2}{4M^2}\right)}$$

$$\to -(\vec{k}^2)\left(\frac{1}{N}f_D^2\right)\left(\frac{1}{\mu^2}\right)^2, \quad \mu^2 = \left(\frac{1}{m+\lambda M}-\frac{\lambda}{M}\right)\frac{12M^3}{\lambda(1-\lambda^2)}$$

(7.1)

Since the residue of the double pole determines the matrix element of the effective Lagrangian at this order, we find a canonical kinetic term for the dilation,

$$L_{eff} = \frac{1}{2}(\partial D)^2\left[1+\Sigma\, a_n D^n\right]$$

(7.2)

The higher order terms in the effective Lagrangian are determined from the higher point correlation functions. Fortunately we are able to calculate these correlators by simply taking the derivative of the two-point correlator with respect to the bare quark mass, m.

$$\langle J_0(\vec{k})J_0(-\vec{k})J_0(0)......J_0(0)\rangle_{n-insertions} = (-\partial_m)^n \langle J_0(\vec{k})J_0(-\vec{k})\rangle$$

(7.3)

where additional current insertions are evaluated at zero momentum transfer. In our expression for the expanded correlator, Eq.(7.1), the mass parameter, m, only enters through the dilaton mass, $\mu^2(m)$, we obtain

$$a_n = (n+1)(\partial_m \mu^2)^n/(f_D)^n$$

(7.4)

The effective Lagrangian in Eq.(7.2) can be resumed to obtain

$$L_{eff} = \frac{1}{2}(\partial D)^2 / \left[1-\left((\partial_m\mu^2)/f_D\right)D\right]^2$$

(7.5)

The constant in the denominator is to be evaluated at the critical coupling for the conformal symmetry with $m = M/\lambda - \lambda M$.

$$C = -\left(\partial_m \mu^2\right)/f_D = \frac{1}{(m+\lambda M)^2} \frac{12M^3}{\lambda(1-\lambda^2)} \frac{1}{f_D} = \frac{12M}{f_D} \frac{\lambda}{1-\lambda^2} \tag{7.6}$$

with $f_D$ as given in Eq.(5.3). For weak coupling, $(0 < \lambda < 1)$, the coefficient, C, is positive while it turns negative for the region, $(1 < \lambda < 2)$. It is clear that the effective Lagrangian in Eq.(7.5) is equivalent to a massless, free action for a canonical scalar field, $\varphi_D$, defined as

$$\varphi_D = \frac{1}{C}\log[1+CD], \quad L_{eff} = \frac{1}{2}(\partial \varphi_D)^2 \tag{7.7}$$

The shift symmetry of the Goldstone phase is evident in this version of the effective Lagrangian.

Using this identification, we obtain a more complete description of the scalar current in the effective field theory. The expression for the scalar current in Eq.(5.3) should be modified to include its vacuum expectation value. We find

$$J_0(x) = \vec{\psi}^+\vec{\psi}(x) = f_D D + \langle J_0(x)\rangle = f_D \frac{1}{C}\exp(C\varphi_D) \tag{7.8}$$

where

$$f_D/C = f_D^2 \left(\frac{1}{\lambda} - \lambda\right)\frac{1}{12M} = \frac{NM^2}{2\pi}\frac{1}{\lambda^2}\left(\frac{1}{\lambda} - \lambda\right) \tag{7.9}$$

Note that Eq.(7.9) is odd under the reflection symmetry, $\lambda \to -\lambda$, as expected from the symmetries associated with the scalar density.

$$J_0(x) = \frac{NM^2}{2\pi}\frac{1}{\lambda^2}\left(\frac{1}{\lambda} - \lambda\right)\exp(C\varphi_D(x)) \tag{7.10}$$

The expressions in Eq.(7.7) and Eq.(7.8) are analogous to the chiral Lagrangian description of pions in quantum chromodynamics in four dimensions. It is interesting to note that scalar current condensate vanishes at $\lambda = 1$ which the transition point between weak and strong gauge coupling. This indicates that the scale symmetry remains unbroken at the transition point and the quark is in the massless symmetric phase of the theory.

**Conclusions.**

I have presented explicit solutions for the massive phase of a U(N) Chern-Simons gauge theory in three dimensions with fermionic matter in fundamental representation of U(N). The theory is scale invariant to leading order in the large N expansion and the existence of a massive solution implies a spontaneous breaking of the scale symmetry. A light scalar bound state emerges in our solution that can be identified as the dilaton, the Goldstone boson associated with the spontaneous breaking of the scale symmetry.

Our solutions for the gap equation for the fermion self-energy and the Schwinger-Dyson equation for the scalar vertex functions are highly nonperturbative, especially in the weak-coupling version of the theory. The massive phase depends on the presence of an extra pole term in the fermion gap equation that is interpreted as arising from a zero mode contribution, presumably an artifact of the light-cone gauge formalism. The solution also requires the introduction of an explicit bare quark mass parameter that is tuned to recover the critical behavior.

Our solution is valid away from the critical point where the dilaton becomes massive. The weak coupling version of the theory develops a hierarchy of mass scales with the dilaton dominating the infrared region and the conformal behavior describing the ultraviolet region and an intermediate region where two-point correlator is constant. We can also observe the behavior of the theory as it approaches the upper limit for the gauge coupling constant at $\lambda = 2$.

We compute the effective action for the dilaton and provide a description of the scalar current condensate. Our results are obtained directly from the low energy limit of the exact correlation functions. These results are analogous to the chiral Lagrangian description of the low energy dynamics of quantum chromodynamics in four dimensions.

Our exact solutions provide a novel laboratory for studying the mechanisms of dynamical symmetry breaking. They may also help in developing an understanding the subtle relations between the various descriptions of the Chern-Simons theories at large N. At the very least, we have solved a very nonperturbative theory that may provide lessons for other quantum field theories. While highly speculative, it would be interesting to explore whether aspects of the dynamical symmetry breaking can be realized in physical applications of Chern-Simons gauge theories.

**Acknowledgements.**

I wish to thank Moshe Moshe for inspiring my interest in the large N Chern-Simons gauge theories and for our joint study of the massive phase of the scalar quark version of the theory. I appreciate the many useful comments and criticisms of my Fermilab colleagues, Estia Eichten, Joe Lykken, Chris Hill and others. I also thank Ofer Aharony for his incisive comments.

**Appendix A - Prefactor Discussion.**

At the end of Section 4 we noted a discrepancy between the results obtained in this paper for the scalar current correlator and previous computations done strictly in the conformal phase of the theory [2,3,4,5]. We investigate the source of this discrepancy and argue that the strict implementation of the conformal symmetry supports the present analysis.

Aside from some different sign conventions, our Eq.(4.5) agrees with the results using calculations done strictly in the conformal phase of the theory except for the prefactor $(1/\lambda - \lambda)$ which was found to be $(1/\lambda)$ in the related work [2,3,4,5,9]. In comparing our detailed calculations with those of G. Gur-Ari and R. Yacoby [5], the results presented here for the vertex equations and the form of the solution for the scalar current correlator are in exact correspondence. The apparent difference arises in their use of an explicit transverse momentum cutoff, analogous to a sharp Fermi surface or band structure, while our solution maintains the scale symmetry of the conformal phase.

In our notation, the scalar current correlator is determined by the high transverse momentum limit of the nonsinglet vertex function as shown in Eq.(4.3).

$$\langle J_0(k_3) J_0(-k_3) \rangle / N = -\frac{1}{2\pi\lambda} \langle ip_- V_+(p,k_3) \rangle, \quad \vec{p}^2 = p_s^2 \to (\infty, \Lambda^2) \tag{A.1}$$

In the conformal phase, the nonsinglet vertex function in Eq.(3.4) takes the form

$$\begin{aligned} ip_- V_+(p,k_3) &= (\lambda p_s - ik_3/2) A(k_3) + B(k_3)(\lambda p_s + ik_3/2)\Phi(p,k_3) \\ &= \lambda p_s V_0(p,k_3) + i(k_3/2)(-A(k_3) + B(k_3)\Phi(p,k_3)) \end{aligned} \tag{A.2}$$

where $\Phi(p,k_3) = \exp(-2i\lambda \arctan(k_3/2p_s))$. The boundary conditions for the A and B coefficients are given in Eq.(3.6) as

$$p_s \to (\infty, \Lambda), \ V_0(p,k_3) \to 1; \quad p_s \to 0, \ ip_- V_+(p,k_3) \to 0 \tag{A.3}$$

The infrared boundary condition is the same for both solutions and gives the relation

$$A(k_3) = B(k_3)\exp(-i\lambda\pi \cdot sign(k_3)) \tag{A.4}$$

The ultraviolet boundary condition on the singlet vertex function, $V_0(p,k_3)$, determines the second constraint on the A and B coefficients. Using our conformal boundary conditions, no new scale is introduced and

$$V_0(p,k_3) \to 1, \ p_s \to \infty \Rightarrow A(k_3) + B(k_3) = 1 \tag{A.5}$$

Note there are no ultraviolet divergences involved in the solution of the integral equations for the vertex function.

Using the explicit transverse momentum cutoff, the vertex functions are slightly modified and the boundary condition becomes

$$V_0(p,k_3) \to 1, \ p_s \to \Lambda \Rightarrow A(k_3) + B(k_3)\exp(-2i\lambda\arctan(k_3/2\Lambda)) \tag{A.6}$$

The asymptotic behavior of the nonsinglet vertex function needed to determine the scalar current correlator is now modified in the two cases. Both solutions have a linear divergence that is independent of the longitudinal momentum, $k_3$.

Using the cutoff solution we find the expansion

$$\begin{aligned} ip_- V_+(p,k_3) &\to \lambda\Lambda V_0(\Lambda,k_3) + i(k_3/2)(-A(k_3) + B(k_3)\Phi(\Lambda,k_3)) \\ &\to \lambda\Lambda - (|k_3|/2)\tan(\lambda\pi/2) \end{aligned} \tag{A.7}$$

Discarding the linear divergence, $\Lambda$, and using the relation of Eq.(4.3),(A.1), we obtain the solution for the scalar current correlator of reference (5) with the prefactor of $(1/\lambda)$.

Using the conformal solution, additional terms are generated from the expansion of the singlet vertex function, $V_0(p,k_3)$.

$$\begin{aligned} ip_- V_+(p,k_3) &\to \lambda p_s V_0(p,k_3) + i(k_3/2)(-A(k_3) + B(k_3)\Phi(p_s,k_3)) \\ &\to \lambda p_s - i\lambda^2(k_3/2) + \lambda^2(|k_3|/2)\tan(\lambda\pi/2) - (|k_3|/2)\tan(\lambda\pi/2) \\ &\to \lambda p_s - i\lambda^2(k_3/2) - (1-\lambda^2)(|k_3|/2)\tan(\lambda\pi/2) \end{aligned} \tag{A.8}$$

Again, discarding the terms linear in $p_s$ and $k_3$, and using Eq.(4.3),(A.1), we obtain our solution for the conformal phase as given in Eq.(4.6).

From the above analysis, it would seem that the effects of introducing a sharp transverse momentum cutoff in a conformal theory are not necessarily removed by taking the cutoff to infinity. This cutoff sensitivity could be an artifact of using light-cone gauge. However, we think the use of the conformal solution is justified and will assume that the results obtained using Eq.(4.5) and Eq.(4.6) are correct.

We remark that the extra term found in (A.8) can also be seen in the computation of the three-point correlator involving a vector current and two scalar currents. At zero momentum transfer for the vector current, this correlator can be obtained from the $k_3$ derivative of the two-point scalar current correlator. This amplitude is finite and seems to require the presentce of the additional term in (A.8) for the two-point correlator to obtain a consistent result.

**References.**

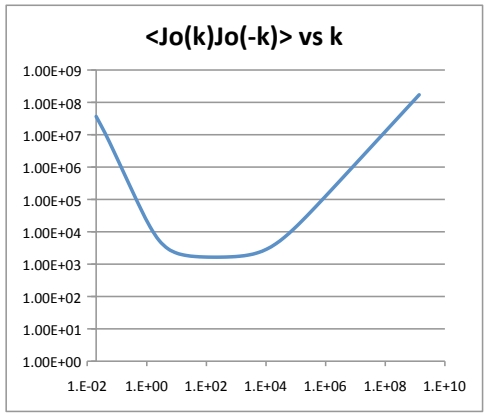
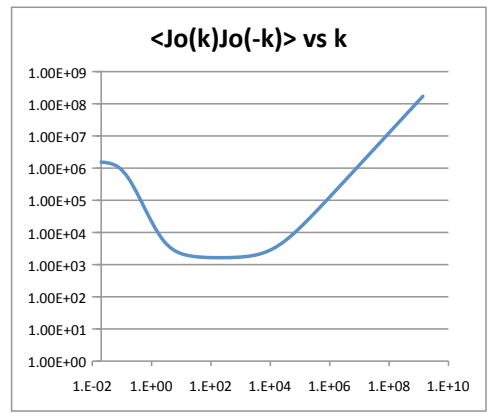

Figure 1.a. λ=0.01, μ=0　　　　　　　　　　Figure 1.b. λ=0.01, μ≠0

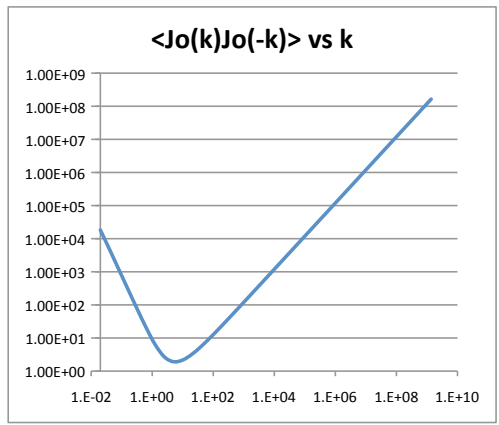
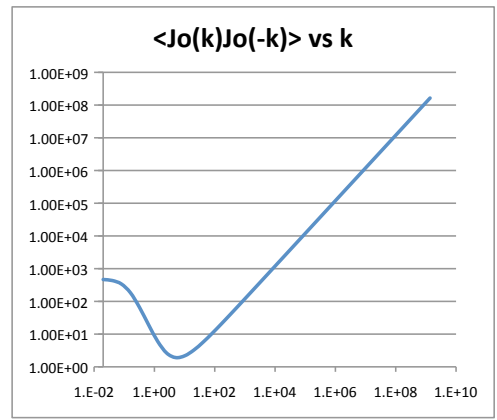

Figure 2.a. λ=0.5, μ=0　　　　　　　　　　Figure 2.b. λ=0.5, μ≠0

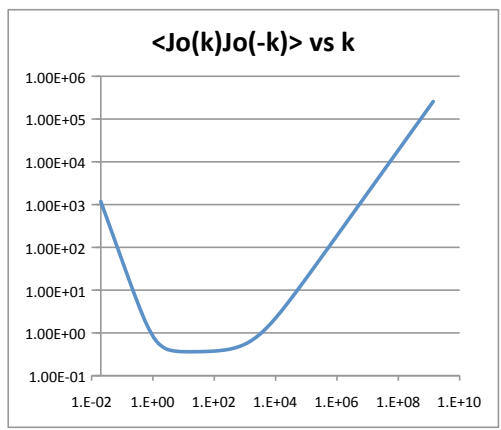
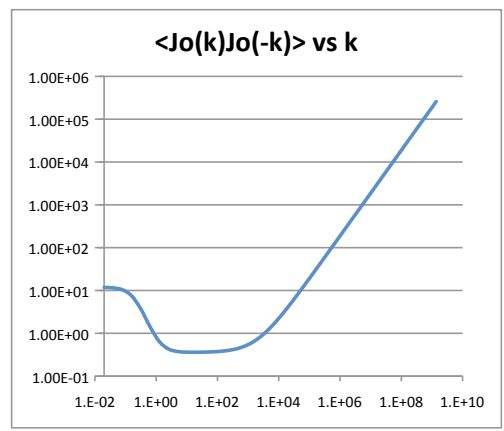

Figure 3.a. λ=1.999, μ=0　　　　　　　　　Figure 3.b. λ=1.999, μ≠0